 \DocumentMetadata{}
\documentclass[sigconf,nonacm=true]{acmart}

\usepackage{graphicx}
\usepackage{subfig}  
\usepackage{balance}  
\usepackage{enumitem}
\usepackage{stfloats}
\usepackage{latexsym} 
\usepackage{url}
\usepackage[textsize=tiny]{todonotes}
\usepackage{xcolor}
\usepackage{tabularx}
\usepackage{booktabs}
\usepackage[ruled,vlined,linesnumbered,noresetcount]{algorithm2e}
\usepackage{varwidth}
\usepackage{tablefootnote}
\usepackage{framed}
\usepackage[bottom]{footmisc}
\usepackage{soul}
\usepackage{eurosym}
\usepackage{titlesec}
\titlespacing{\paragraph}{5pt}{2pt}{2pt}
\titlespacing{\subsection}{1pt}{5pt}{2pt}
\titlespacing{\subsubsection}{1pt}{5pt}{2pt}

\begin{document}
\title{
    Bringing Data Transformations Near-Memory for Low-Latency Analytics in HTAP Environments
}

\author{Arthur Bernhardt$^{*}$, David Volz$^{\#}$, Sajjad Tamimi$^{\#}$,   Andreas Koch$^{\#}$, Ilia Petrov$^{*}$}
\affiliation{%
	\institution{\textsuperscript{\#}Embedded Systems and Applications Group, $^{*}$Data Management Lab}
	\institution{$^{\#}$TU Darmstadt, $^{*}$Reutlingen University}
	\country{Germany}
}

\renewcommand{\sf}{\sffamily}
\newcommand{\tn}{tn}
\newcommand{\tp}{tp}
\newcommand{\fp}{f\!p}
\newcommand{\fpr}{f\!pr}
\newcommand{\pot}{pot}
\newcommand{\nKVtitle}{\textsf{\textbf{\lowercase{\underline{n}}KV}}}
\newcommand{\nKVcaption}{\textsf{\textbf{nKV}}}
\newcommand{\nKV}{{\sf \lowercase{\underline{n}}KV}}
\newcommand{\neoDB}{{\sf \lowercase{neo}DBMS}}
\newcommand{\NDT}{{\sf nDT}}

\pagestyle{plain}

\begin{abstract}
In this paper we propose an approach for executing data transformations near- or in-storage on intelligent storage systems. The currently prevailing approach of extracting the data and then transforming it to a target format suffers degraded performance during transformation and causes heavy data movement. Our results show robust performance of foreground workloads and lower resource contention. Our vision draws architectural opportunities in multi-engine and multi-system settings, as well as for reuse.
\end{abstract}

\maketitle

\section{Introduction}
\label{sect:intro}

\paragraph{Motivation.}
Data analytics pipelines are widely used to extract insights from OLTP or data ingestion systems or answer prediction queries. To this end, data is exported and transformed into open columnar data formats \cite{Hunagchen:comparison:arxiv:2023} such as Arrow \cite{Arrow}, ORC \cite{ORC} or Parquet \cite{Parquet}, to be consumed by analytical frameworks for analytics and machine learning. However, exporting and transforming OLTP data is slow and expensive \cite{Muehleisen:WireProtocol:VLDB:2017,Pavlo:MainliningDBMS:VLDB:2020}, as it is resource intensive and incurs significant data movement. The latter  is the root cause for many inefficiencies \cite{Manos:VisionRowCol:SigmodRec:2021,Dally:ITC} as OLTP data must be scanned and extracted, incurring slow I/O, and memory and CPU contention with foreground workloads. 
Given the trend of super-linear data growth rates, the impact of movement will continue to increase in the future. 

What's more, data analytics and machine learning pipelines are deployed on popular Python frameworks such as Pandas, PyTorch, TensorFlow, or SciKit-Learn. Noticeably, such wide-spread analytics and machine learning tools do not integrate with DBMS beyond bulk dataset transfers \cite{Muehleisen:WireProtocol:VLDB:2017}. To reduce the export and transformation cost, typically small samples are extracted, which inevitably reduces the accuracy of analytics algorithms \cite{Muehleisen:WireProtocol:VLDB:2017}.
As these are separate tools, the advantages of modern Hybrid Transactional Analytical Processing (HTAP) DBMS for analytics pipelines cannot be exploited. 
The pressure to export, transform and import large amounts of data across systems is likely to increase, since Python tool eco-systems and DBMS are separate systems.

\paragraph{State-of-the-Art Overview.}
Given the strong focus on data analytics and the necessity to export data from various sources, there is a rapid spread of open formats (e.g., Arrow \cite{Arrow}, ORC \cite{ORC} or Parquet \cite{Parquet}) and analytical systems that support them \cite{velox,duckdb}. 
Yet, prior work has shown that exporting and transforming data to the target format is a major source of inefficiency \cite{Muehleisen:WireProtocol:VLDB:2017,Pavlo:MainliningDBMS:VLDB:2020} (Fig. 1, top left). To this end, new designs \cite{Pavlo:MainliningDBMS:VLDB:2020} (Fig. 1, top middle) investigate storing the main part of the dataset in Arrow, with a novel OLTP engine on top. DuckDB \cite{duckdb} or Velox \cite{velox} among other systems can natively operate on Arrow.

However, the majority of the proposed approaches, while efficient and universal are hardware-oblivious and do not primarily address the data movement problem. Yet, there has been significant progress in the field of hardware, especially intelligent-storage or -memories. With the advance of semiconductor memories and fabrication processes, it has become economical to produce combinations of storage and compute elements packaged on the same device. Such intelligent storage devices allow offloading operations for on-device processing and offer better device-internal I/O characteristics (bandwidth, latencies) than device-to-host. Hence, they offer a possibility for data transformations and addressing data movement.
\cite{Manos:VisionRowCol:SigmodRec:2021} a seminal vision for transparent data transformations that can be offloaded to smart memories. Along the same lines \cite{Manos:RelationalMemory:EDBT:2023} demonstrate an approach for near main-memory row-column transformations.

\begin{figure}[t]
	\centering	\includegraphics[width=\linewidth]{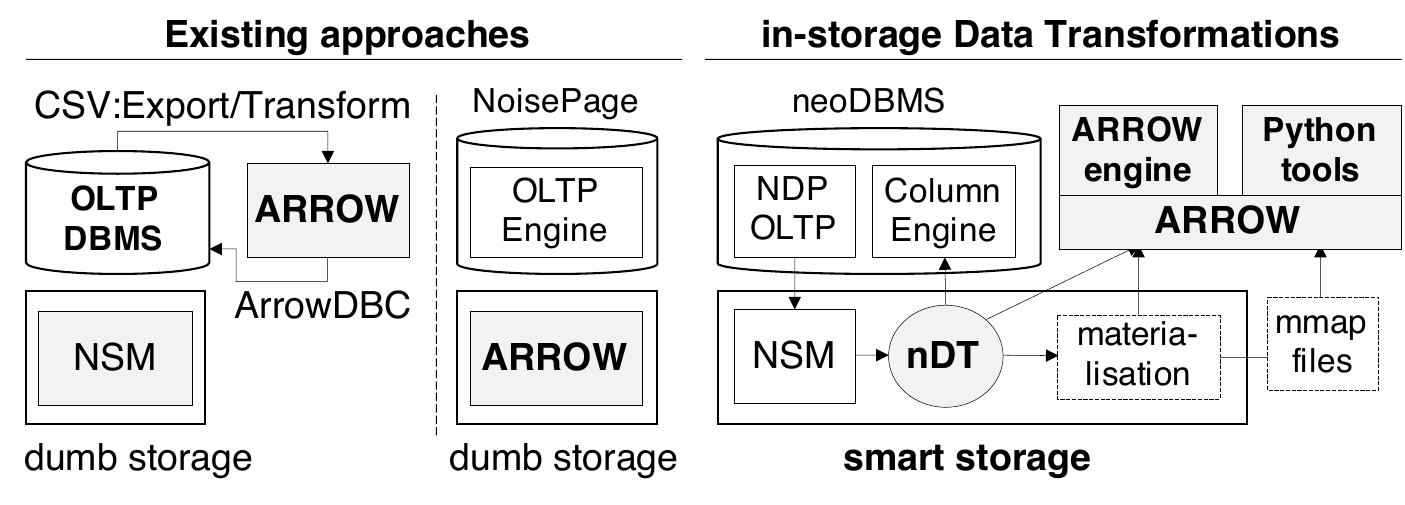}
		\centering	\includegraphics[width=\linewidth]{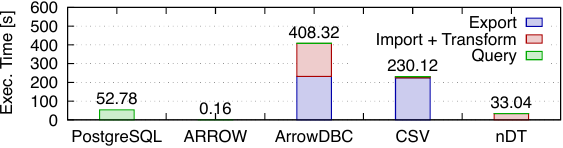}
	\caption{(Top) various alternatives for data transformations: by  exporting DBMS data and importing it in the target system; directly accessing the source DBMS; and near-storage. \\ (Bottom) performance comparison of these approaches under HTAP settings with simultaneous OLTP workload.}	
\vspace{-15pt}
	\label{fig:headpict}
\end{figure}

\paragraph{Contributions.}
In this paper we propose an approach that aims at offloading data transformations to smart storage. We argue that \NDT{} can be executed as near-data processing (NDP) operations and are a natural fit for smart storage to reduce data movement.

The main idea is to perform data layout transformations (row-to- column/Arrow) on-device on compute cores close to the physical data location, instead of effortfully moving all the OLTP data to the host (Fig. 1, top right). Thus, \NDT{} benefit from the high on-device bandwidth. What's more, \NDT{} are executed in an asynchronous and interruption-free manner on-device without any intervention with the host-DBMS or contention for host resources, e.g. buffers. 

A key aspect of \NDT{} is that in HTAP settings, they execute against a transactionally consistent snapshot of the database computed on-device. They offer snapshot isolation transactional guarantees and high data freshness, thus enabling multi-engine settings. \NDT{}  also offer time-travel semantics if requested by the application. 

Another major aspect is that \NDT{} promote co-execution and yield robust performance. In HTAP settings the host-DBMS continues serving the foreground workload, while \NDT{} execute interruption- and contention-free on device.

Furthermore, \NDT{} allow the streaming of their results to the host for multi-engine settings, but also materializing the results on device for reuse or for supporting multi-system settings. Notably, \NDT{} allow for fast incremental updates of materialized transformed data (delta-\NDT{}), i.e., given a prior materialization, only the fresh delta can be computed, in a repeated transformation run.

Lastly, \NDT{} are economical. Transformations are  faster in-memory, but also more expensive, given the costs of $\sim$20000\$/TB server DDR4 versus $\sim$500-1000\$/TB of high-performance smart flash storage.

The potential of the proposed in-storage transformations is explored in Figure \ref{fig:headpict}. We compare the performance of the different approaches\footnote{Unfortunately, we could not compare to NoisePage \cite{NoisePage}, due to compilation issues on our ARM Neoverse N1-SDP \cite{N1SDP}.} under HTAP settings and the CH-benchmark \cite{Cole:CH-Benchmark:DBTEST:2011} (SF 100), where we execute Q6 in parallel to the OLTP load. As baselines we consider PostgreSQL, and the Arrow engine \cite{ACERO} that processes Q6 only after the transformation. The execution time under PostgreSQL indicates the overhead of data movement, while the execution times of other systems indicate the additional overheads of  data export and transformation. \NDT{} is fast as it mitigates data movement by offloading the Arrow transformation to storage.

\begin{figure}[b]
	\begin{center}
        \includegraphics[width=\columnwidth]{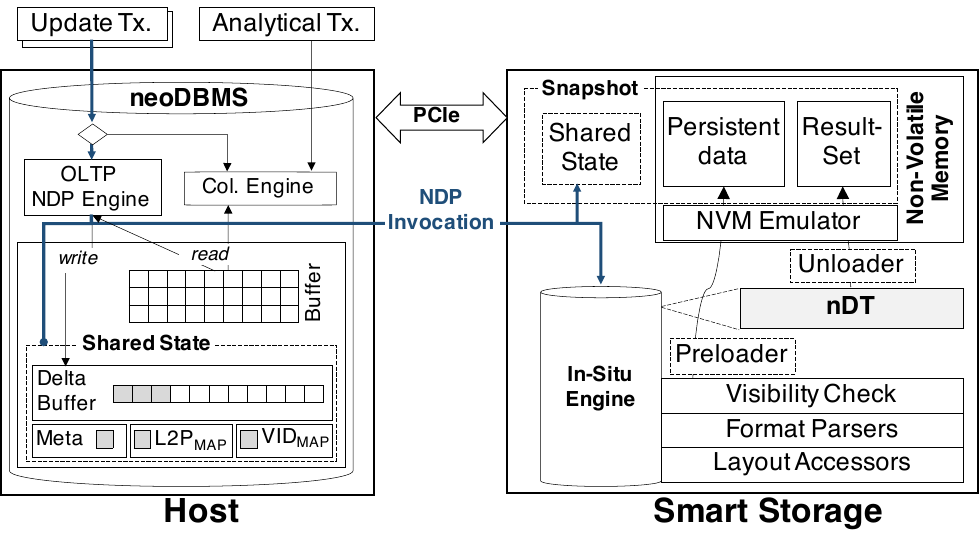}  
		\caption{\neoDB{} NDP architecture and \NDT{} integration.}
	\label{fig:neoDB:arch}
 \vspace{-10pt}
	\end{center} 
\end{figure} 

\paragraph{Outline.} 
The paper is organized as follows. We first describe DBMS design decisions for necessary for \NDT{} and overview the execution model. Sect. \ref{sect:multi:engine} and \ref{sect:multi:system} describe opportunities offered by \NDT{} in multi-engine or multi-system settings, and present initial experimental results in our NDP-DBMS \neoDB{}. We conclude in Sect. \ref{sect:conclusions} and discuss future directions.

\section{DBMS design for in-storage  transformations} 
\label{sect:transforms}
A key idea of this paper is that \NDT{} can be executed as a near-data processing (NDP) operation on smart storage, mitigating some of the well-known overheads that the layout dichotomy and transformation pose on DBMS designs \cite{Pavlo:Archipelago:SIGMOD:2016}. To this end, an NDP-DBMS (such as PolarDB \cite{PolarDB}, \neoDB{} \cite{Vincon:UpdateAwareNDP:vldb:2022} or \nKV{} \cite{Vincon:nKV:DAMON:2020,Vincon:nKVinAction:VLDB:2020}) capable of managing the execution is needed. In the following, we overview the design of such NDP-DBMS for \NDT{} on the basis of \neoDB{} \cite{Vincon:UpdateAwareNDP:vldb:2022} and describe the \NDT{} execution model.

\paragraph{Design Overview.}
As the foreground workload is executed, \neoDB{} incrementally accumulates the modifications, and places them in a small shadow area called \emph{shared-state} \cite{Vincon:UpdateAwareNDP:vldb:2022} (Fig. \ref{fig:neoDB:arch}). The shared-state is regularly propagated at low overhead as it is kept small and configurable in the range of a few hundred KB to a few MB. Notably, the shared-state is the only delta between the {\em working set} in the large host-DBMS main-memory with the most up-to-date data, and the much larger but colder and {\em complete} dataset on smart storage.

A key point is, that after a shared-state propagation, the smart storage attains all data necessary to construct a transactionally consistent DBMS snapshot in-situ. The on-device processing begins from the shared-state and only then moves to the cold data, as version records in the latter may have been invalidated by newer ones in the former. The shared-state is propagated in two distinct modes: (a) regularly, whenever it reaches a predefined size; and (b) alongside an NDP/\NDT{} invocation where it gets snapshotted and combined with a list of transactions currently in-flight.

The Delta-Buffer (Fig. \ref{fig:neoDB:arch}) is a key element of the shared-state. It accumulates the versions newly created by active transactions, 
while predecessor versions remain in the main buffer. Furthermore, the shared-state accumulates incremental changes to auxiliary structures such as data dictionary definitions, the address-mapping table ($L2P_{map}$), or the VID table ($VID_{map}$), which we introduce below. 

For one, to support shared-state propagation and on-device  snapshot construction, \neoDB{} organizes version records as a singly backwards-linked \textit{new-to-old (N2O)} list \cite{Vincon:UpdateAwareNDP:vldb:2022,Bernhard:neoDB:ICDE:2022}. To mark the entry-point of a chain, \neoDB{} employs a $VID_{map}$ containing the RecordID of the latest version record for each tuple. All version records in a chain have the same \emph{Virtual ID} (VID) as they belong to the same tuple. For another, to enable in-situ RecordID/address-resolution and navigation, \neoDB{} maintains a logical to physical address-mapping ($L2P_{map}$) that is incrementally propagated with the shared-state.

\subsection{Overview of the Execution Model}
\label{sect:exec}

\subsubsection{NDP/nDT Invocation.}
Before offloading any NDP operation, e.g., \NDT{} to smart storage, the NDP-DBMS prepares the execution. It retrieves the unique \texttt{TxID} of the calling transaction and creates a list of all transactions currently in-flight, both of which are necessary for in-situ execution and in-situ snapshot creation. Next, an \NDT{} invocation is constructed, which comprises the transaction context, the involved DB objects, and the operations to be executed. Along these lines, \neoDB{} snapshots the current shared-state, to be propagated alongside the NDP call.

During the preparation, the NDP-DBMS estimates and allocates the storage and on-device resources necessary for the invocation. The estimation is performed based on the expected selectivity, the necessary compute resources or cache/FPGA-BRAM sizes, and the current device load.
Storage space is provided as new DB pages, reserved and/or allocated for the pending \NDT{} invocation. The NDP space management is performed solely by the DBMS. Therefore, if the \NDT{} execution runs out of storage, it requests additional pages from the DBMS at the cost of an extra roundtrip. To this end, the current \NDT{} execution is temporarily suspended and only resumes after the space allocation completes. Noticeably, \neoDB{} configures and allocates on-device resources for each NDP call. 

Once the \NDT{} invocation is performed, the worker threads of the NDP engine get suspended and sleep-wait for the response of the in-situ engine. Noticeably, the NDP engine can perform further invocations, which will execute against their own on-device snapshots.

\subsubsection{On-device nDT Scheduling/Preparation. }
\NDT{} processing on the individual smart storage processing elements (PEs) begins with the arrival of the \NDT{} invocation command. The \textit{in-situ engine} schedules the workload across the number of PEs specified in the invocation. This is done by partitioning the in-situ $VID_{map}$ uniformly over all PEs in a round-robin manner and assigning each partition to a PE. The pre-allocated storage for the \NDT{} invocation is uniformly distributed over and assigned to all PEs.

Next, the in-situ engine launches the \NDT{} (Sect. \ref{sect:ndt}) as partitioned \textit{NDP jobs} across the number of PEs specified in the invocation. The jobs start by performing a visibility check (Sect. \ref{sect:vis:check}) for each tuple starting from the latest version record (referenced by $VID_{map}$ entry) in their partition. To this end, \NDT{} processing must be able to navigate on-device, which we briefly explain next.

\subsubsection{In-situ Navigation and Data Interpretation.}
\label{sect:data:interpret}
\NDT{}, must access, interpret and navigate through the persistent binary data in-situ without costly host interaction and transfer-intensive roundtrips. To this end, schema information is propagated with the \NDT{} invocation to the device. It comprises information about DB-objects, their schema (attributes, types, sizes), or physical representation. The on-device NDP infrastructure employs that information to support data layout accessors for in-situ navigation, and format parsers. These are pre-compiled MicroBlaze \cite{MicroBlaze:2022} binaries.

\textit{Layout accessors} exist for components of the persistent data layout to navigate through the binary data organization in-situ. For example, for the NSM-layout, individual records or their headers.

\textit{Format parsers} extract persistent binary elements (records, values), interpret them semantically, and allow for further processing, mathematical operations, or comparisons. We  distinguish field, record, and page formats and layouts for this purpose.

\paragraph{In-situ address resolution}
is an important part of in-situ navigation necessary to reduce data transfers. It occurs, for instance, during a version-chain traversal, when to retrieve a version-record, the in-situ engine resolves its \textit{RecordID} to retrieve its physical location. To this end, the NDP-DBMS  utilizes the logical-to-physical address mapping ($L2P_{map}$)  that is incrementally propagated alongside the shared-state. It is then used to resolve logical addresses to physical locations without slow and blocking  host-roundtrips.

\subsubsection{In-situ Snapshot Construction.}
\label{sect:vis:check}
To create a snapshot on-device, the visibility check must determine the latest version record of a version chain, committed prior to \NDT{} invocation. 
To this end, the in-situ engine takes the snapshot information from the \NDT{} invocation, i.e. the transaction ID (\texttt{TxID}) of the calling transaction, the shared-state (delta-buffer and the $L2P_{map}$ and the $VID_{map}$).

Given the \textit{N2O} version organization, the visibility check is performed by traversing the records in a version-chain backwards, and comparing their creation timestamps against the \texttt{TxID}. To this end \neoDB{} relies on the byte-addressable nature of the NVM. The visibility task on each PE extracts the entry-point RecordID of each entry (8B transfer) in the PE's $VID_{map}$ partition and resolves it in-situ. This \textit{RecordID resolution} is based on the $L2P_{map}$ (4B transfer) and yields the physical page pointer and a slot offset. It is then passed to the \textit{layout accessor} (Sect. \ref{sect:data:interpret}) on the PE, which retrieves the corresponding NSM page slot (4B transfer). In a successive 4B transfer, the \textit{layout accessor} retrieves the record header and passes it to the format parser to retrieve the transaction timestamp and compare it to \texttt{TxID}. The invalidation timestamp is available from the predecessor (N2O org.) that has already been processed. The visibility decision is taken based on both timestamps. All visible versions are passed on to the \NDT{} execution. 

\begin{figure}[b]
	\begin{center}
        \includegraphics[width=\columnwidth]{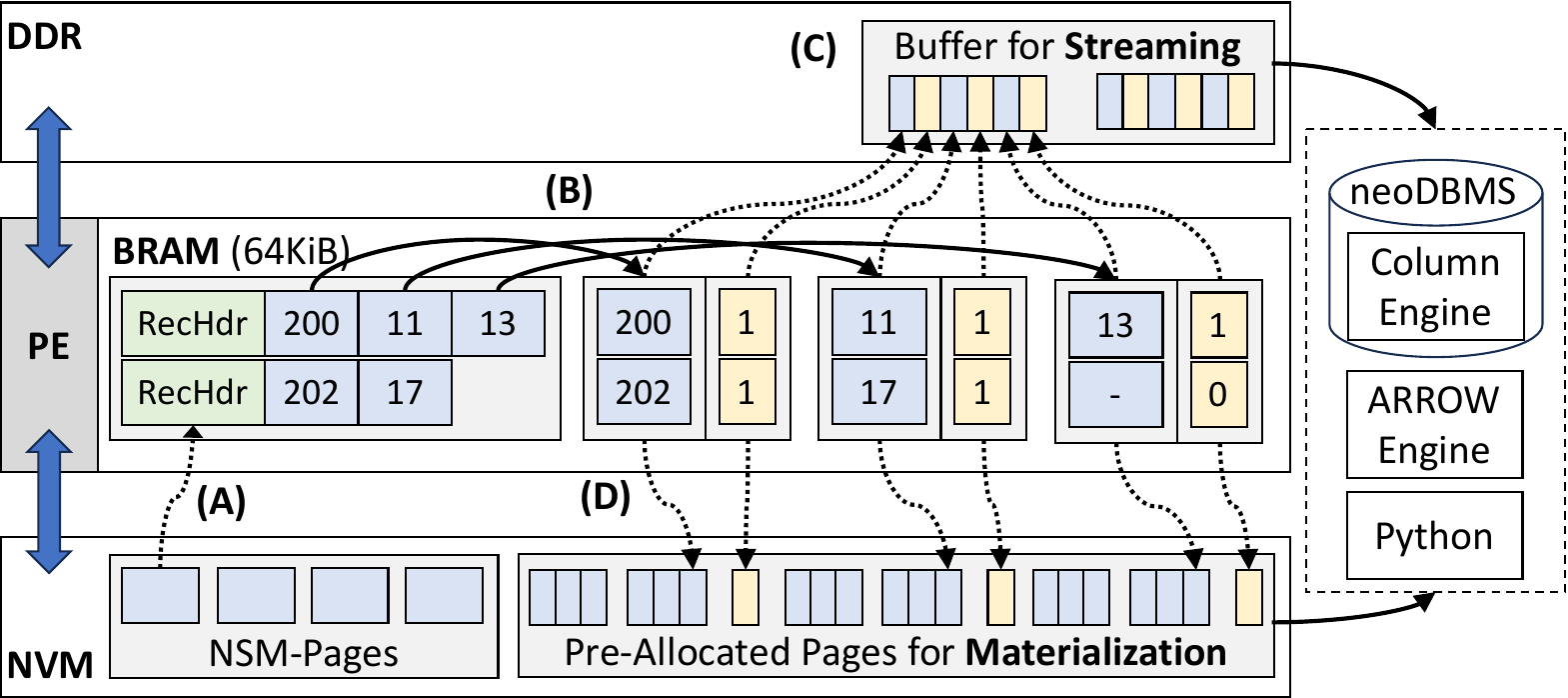}  
		\caption{\NDT{} execution.}
	\label{fig:neoDB:ndt}
	\end{center} 
\end{figure}

\subsubsection{In-storage Data Transformations.}
\label{sect:ndt}
\NDT{} allow in-storage transformations of NSM-formatted (row-organized) OLTP-data to an Arrow compatible columnar storage format. The Arrow format serves as a cross-language, cross-platform data layer that enables seamless data interchange between various systems and programming languages. Arrow employs a memory layout that adheres to a zero-copy approach, optimizing data movement and minimizing overhead. Metadata (Arrow Schema) provide the blueprint for interpreting and processing the data without unnecessary conversions. Moreover, Arrow introduces record batches, a table-like data structure, which represent a logical grouping of data across columns.

To transform the NSM data to Arrow, the smart storage processing elements (PEs) employ BRAM (64 KiB) as fast scratchpad memory (Fig. \ref{fig:neoDB:ndt}). BRAM provides a high-performance memory resource that facilitates quick data access and manipulation. The in-situ engine partitions the available BRAM over the PEs involved in the \NDT{} based on the schema and projection information from the \NDT{} invocation. To this end, 8KiB BRAM partition is reserved and used for loading NSM records after snapshot computation and visibility checks (Fig. \ref{fig:neoDB:ndt}.A). The remaining free BRAM space is partitioned equally for each attribute taking part in the transformation and requested by the projection. For each attribute that may contain NULL values, an additional validity bitmap partition is created. Moreover, variable-length attributes result in an extra BRAM partition to store offset information. Our experiments indicate that BRAM can be sized much smaller without compromising performance. For instance, using only 16KiB BRAM results in the same transformation performance of the CH/TPC-C \texttt{Orderline}  table compared to 64KiB.

The row-to-columnar layout transformation relies on BRAM to BRAM copies for efficient data rearrangement (Fig. \ref{fig:neoDB:ndt}.B). Layout accessors would be sufficient for fixed-length data types during the transformation process, as they also account for NSM layout alignment rules in data types. However, to handle variable-length data types and type conversions like PostgreSQL timestamps to UNIX epoch, format parsers are employed to handle the specific transformation. The record header is evaluated to determine NULL values in any attribute and fill the validity bitmap partitions accordingly. When BRAM partitions become full, they are flushed to assigned pages in either DDR (Fig. \ref{fig:neoDB:ndt}.C) or NVM (Fig. \ref{fig:neoDB:ndt}.D).

\subsubsection{\NDT{} On-Device Result Handling.}
\label{sect:results:handling}
\NDT{} supports different result consumption modes (Fig. \ref{fig:neoDB:ndt}), which address specific transformation requirements and help optimize the data processing pipeline.

\paragraph{Streaming.}
\NDT{} allow the streaming of results up to different host systems as they are produced while transformation proceeds. This mode leverages the concept of Arrow streams, and the transient nature of Arrow. \NDT{} ensures efficient interleaving of result production and consumption by the host engine. To this end, multiple small buffers in on-device DDR are reserved to host transformed results. Whenever they get full, a host notification is triggered and a new result batch can be loaded. The PEs continue to concurrently prepare the next batch of data, ensuring a continuous data flow.

\paragraph{On-device materialization.}
More importantly, the in-situ engine may also materialize the transformation results on device.
In materialization mode, the PEs flush their BRAM partitions into pre-allocated NVM pages that are uniformly distributed across all PEs, allowing for better parallelism. Upon \NDT{} completion, PEs return the addresses and the sizes of every columnar structure and their validity bitmaps and offset vectors, while unused space is marked free. In the rare case of insufficient result space, the in-situ engine temporarily suspends the \NDT{} jobs and requests more free space from the NDP-DBMS.

\begin{figure}[b]
	\begin{center}
        \includegraphics[width=\columnwidth]{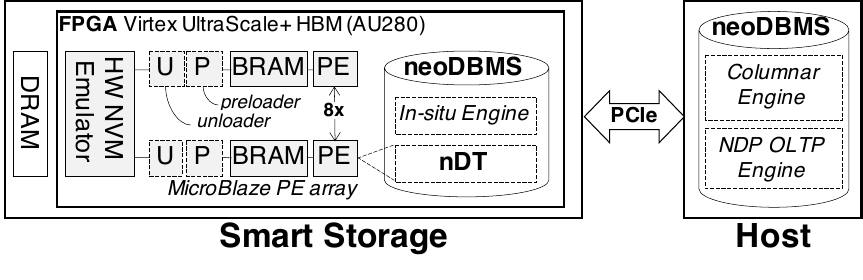}  
		\caption{\neoDB{} hardware architecture.}
	\label{fig:neoDB:HW}
	\end{center} 
\end{figure}

\subsection{Hardware Architecture} 
\label{sect:arch:HW}
We now overview the hardware NDP architecture on smart storage. \neoDB{} deploys \NDT{}, visibility-checker, layout accessors, and format parsers as small pre-compiled binaries loaded at point of \NDT{} invocation into individual processing elements (PE). An array of up to 8 PEs can be used for distributing and executing operations (Fig. \ref{fig:neoDB:HW}). All PEs represent a MicroBlaze \cite{MicroBlaze:2022} scalar soft-core, running at 200MHz and have access to the on-board FPGA memory. A key intuition is that the on-device engine and \NDT{} run as flexible \textit{software-NDP} tasks, programmable in C. The on-device FPGAs have customizable memory hierarchies that can be efficiently utilized for NDP.
Thus, fast on-FPGA BRAM memory (1 clock cycle) is attached to each PE as local scratchpad. Each BRAM is attached to his own \textit{preloader} and \textit{unloader} HW-module, allowing byte-addressable DMA transfers to DRAM. To emulate and match the characteristics of NVM storage, a hardware NVM emulator \cite{Tamimi:NVMulator:ARC:2023} is placed in front of the memory controller to achieve configurable NVM latencies on top of DRAM.

\subsection{Key Challenges} 
\label{sect:challenges}
There are different challenges to the efficient execution of \NDT{}, such as ensuring transactional guarantees for \NDT{} under HTAP settings, or materializing results on-device, and ensuring reuse. In the following, we discuss our ideas for addressing these challenges. 

\paragraph{Transactional guarantees.}
As described above, a key aspect is the ability to perform \NDT{} with transactional guarantees, which is necessary for utilizing \NDT{} in DBMS and for transactional multi-engine processing.  Depending on the requirements of the application, the \neoDB{} transaction manager schedules \NDT{} as part of a multi-engine transaction, involving the NDP OLTP engine and potentially a columnar/analytical engine. To this end, the NDP-DBMS propagates the newest shared-state as part of the NDP call. At this stage, the in-situ engine attains the complete information to construct a transactionally consistent snapshot of the whole database on-device. Now, the row-to-column/Arrow \NDT{} is executed on that snapshot in a streaming manner, keeping the data movement device-internal, while the columnar results are streamed up to the analytical engine.

A second key aspect is the ability to execute \NDT{} asynchronously in an interruption-free manner. The main observation is that slow and blocking host-device roundtrips can be avoided, for instance for schema- or address-information. This way co-execution is achieved, i.e. while the host-DBMS continues processing the foreground workload, the device can independently execute \NDT{} without resource contention or interruption.

\paragraph{On-device materialization.}
A major aspect of \NDT{} is the ability to materialize the transformation results on-device, besides being able to stream them up to the host on-the-fly. This way a richer set of execution models becomes possible, involving for instance vectorized or materialized execution, depending on the operations, the involved engines, or the on-device hardware, e.g. vector PEs. Materialization is economical on smart storage because of the lower price/TB ratio ($\geq$20$\times$) compared to DRAM, or re-computation savings.

Reuse is yet another aspect that materialization enables. For instance, it may allow cooperative  execution of analytics pipelines that share common input or intermediary data. Furthermore, its materialization fosters multi-engine settings.  

Lastly, if an older \NDT{} result is already materialized, a fast incremental/$\Delta$-\NDT{}, should bring it up-to-date at low overhead.

\begin{figure}[t]
	\begin{center}
        \includegraphics[width=\columnwidth]{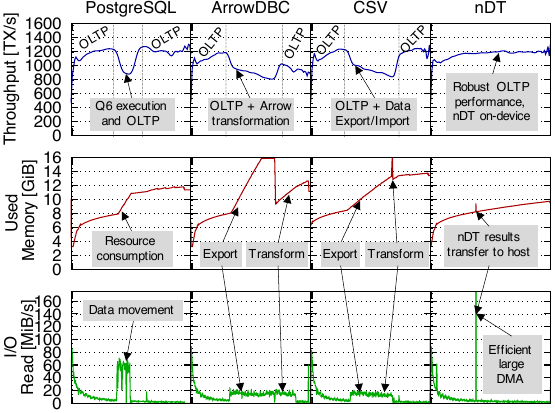}  
		\caption{ Impact of transformations on foreground workload.}
	\label{fig:neoDB:exp_stable_htap}
	\end{center} 
\end{figure}

\section{Opportunities for Multi-Engine Designs.}
\label{sect:multi:engine}
As described earlier \NDT{} offer transactional guarantees. The first part of our vision draws the opportunities for \NDT{} in the context of multi-engine architectures to achieve high data freshness and efficient analytics. \NDT{} suitability is demonstrated in an opening experiment (Fig. \ref{fig:neoDB:exp_stable_htap}), where we execute an HTAP workload, based on the CH-benchmark \cite{Cole:CH-Benchmark:DBTEST:2011} (scale factor 100). The OLAP part is Q6 executed in PostgreSQL as the baseline. (Q6 on Arrow, preceded by data export and transformation was already discussed in Fig. \ref{fig:headpict}).

Under HTAP settings (Fig. \ref{fig:neoDB:exp_stable_htap}), we observe a noticeable impact of the data export and transformation on the foreground OLTP throughput, due to the heavy data movement caused by the export, and the high resource competition (e.g., CPU, memory) during transformation. Under \neoDB{} with \NDT{} Q6 runs on the columnar engine. Due to its asynchronous interruption-free on-device execution, \NDT{} mitigates host resource contention and mitigates data movement. Both result in better (Fig. \ref{fig:headpict}) and robust performance (Fig. \ref{fig:neoDB:exp_stable_htap}) as \NDT{} does not impact the foreground OLTP workload.

\noindent\underline{\textsf{Insight:}} 
\NDT{} may lead to efficient multi-engine DBMS designs by reducing data movement and resource contention. We envision a new type of poly-model stores where different \NDT{}s (e.g., row-column, or KV-to-column), are offloaded to smart storage, re-layout persistent data in multiple data layouts on the way up to the host.

\section{Opportunities for Multi-System Architectures}
\label{sect:multi:system}
The export and transformation overhead is especially prominent in multi-system settings, where OLTP data is exported and transformed to an open format, e.g., Arrow, to be imported into a different system, e.g. SciKit-Learn. A second key part of our vision is to expose \NDT{} and smart storage to different systems, offering further opportunities. 

\paragraph{Interface.}
\NDT{} introduces a standardized connection to DBMS and SQL Interface called NDBC. It mitigates wire-protocol overhead \cite{Muehleisen:WireProtocol:VLDB:2017,Pavlo:MainliningDBMS:VLDB:2020} (Fig. \ref{fig:headpict}), by employing shared memory to provide access to multiple engines/systems. Moreover, NDBC is zero-copy and avoids multiple page copies in DBMS result-set and Arrow. NDBC also allows setting the result-set format, by configuring \NDT{}.

\paragraph{Materialization.}
As described earlier (Sect. \ref{sect:results:handling}), \NDT{} can stream Arrow results to the host as the transformation proceeds, or materialize them on-device, depending on the requirements of the different engines. 

Materialization is economical and low-overhead. Noticeably, smart storage space is cheaper than DRAM, with $\sim$500-1000 \$/TB of high-performance smart flash storage vs.  $\sim$20000\$/TB server-grade DDR4. Moreover, writing out temporary materialized results imposes low overhead as it leverages the high on-device bandwidth ($\sim$16/30 GiB/s vs. $\sim$6,4/12 GiB/s device-to-host). Given the appropriate DBMS space-management strategy, occupied storage is simply scheduled for asynchronous garbage collection, if not needed.

Materialization and streaming enable different execution modes, e.g., materialized and vectorized. An important observation is that with such models, different engines types can be served, e.g., OLTP, OLAP (vectorization), and machine learning (materialization).  

Lastly, materialization fosters easy integration across tool ecosystems. To this end, the \NDT{} results  (Arrow-formatted, OLTP-data) materialized on-device can be exposed to third-party tools as memory-mapped (\texttt{mmap}) \cite{crotty22-mmap} files (Fig. \ref{fig:headpict}, right). As a result, they can be readily consumed by various tools, e.g., SciKit-Learn.

\noindent\underline{\textsf{Insight:}} 
\NDT{} allows efficient consumption of Arrow-transformed OLTP data in multi-system settings. The efficient materialization of \NDT{}-results on smart storage emerges as a key mechanism. Furthermore, low-overhead interfaces such as NDBC and \texttt{mmap}-files support efficient integration in different eco-systems.

\section{Opportunities for Reuse}
\label{sect:reuse}
Noticeably, on-device materialization of \NDT{} results offers major opportunities for reuse. For instance, for cooperatively deployed data analytics and machine learning pipelines \cite{Kaoudi:ReusePipelines:SIGMOD:2022} on the same dataset. 
Reuse is also relevant for iterative analytics (e.g. k-means clustering) over the Arrow-transformed \NDT{}-results. In addition, reuse, foresters materializing statistics computed in previous pipeline stages or different pipelines. Reuse also allows for data sharing across tools. Ultimately, reuse opens an economical tradeoff, as it allows reducing (re-)computation and increasing performance at the cost of more space.

\begin{figure}[b]
	\begin{center}
        \includegraphics[width=\columnwidth]{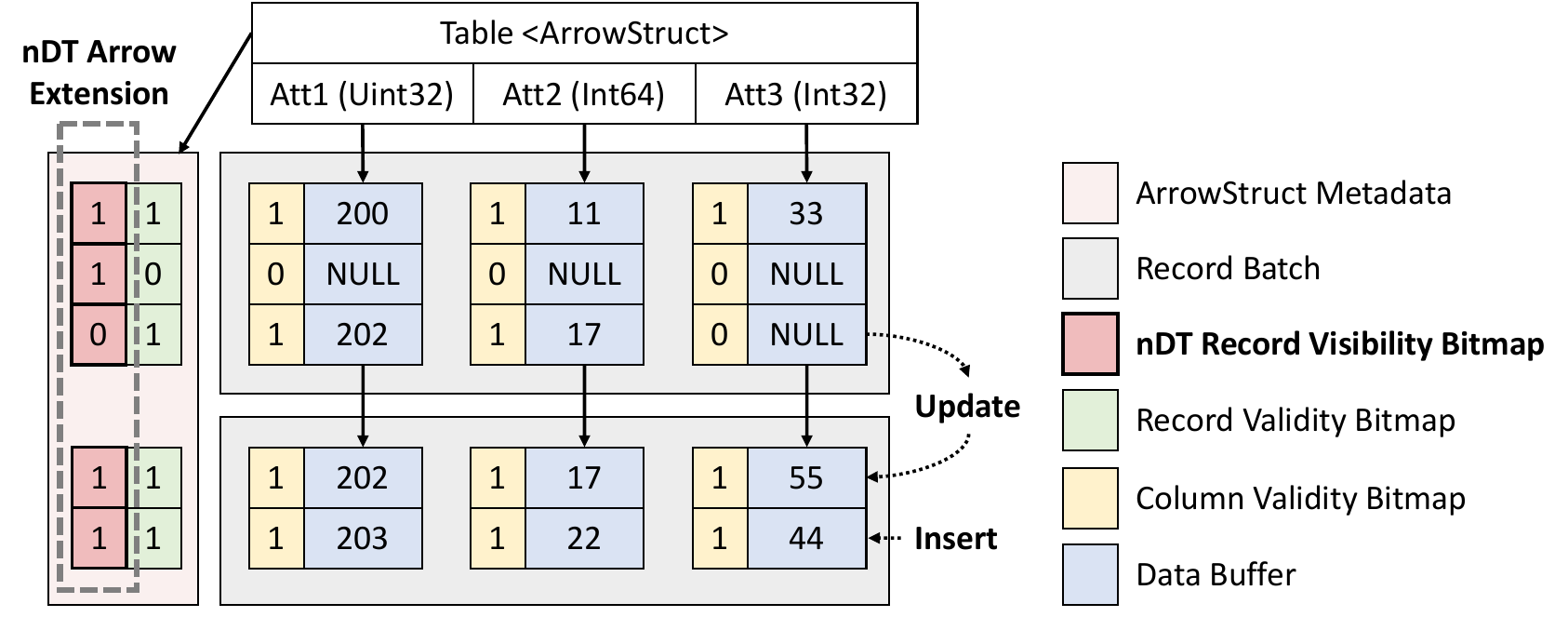}  
		\caption{Integration of delta transformations in Arrow.}
	\label{fig:future:deltaarrow}
	\end{center} 
\end{figure}

A core challenge for reuse, is to avoid the full cost of re-materiali-zation if the original OLTP data has changed. To this end, we envision delta-\NDT{} and an extension to the Arrow format. $\Delta$-\NDT{} efficiently determine the OLTP modifications to the OLTP database that occurred past the initial materialization. The modifications are then transformed to Arrow and appended to the existing materialization. To distinguish between outdated and new Arrow entries, we envision adding a visibility positional bitmap vector (Fig. \ref{fig:future:deltaarrow}).

\begin{figure}[t]
	\begin{center}
        \includegraphics[width=\columnwidth]{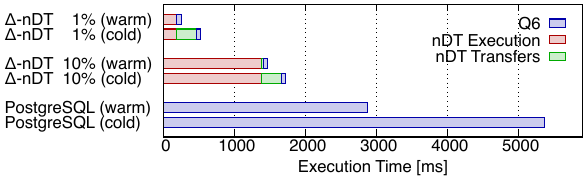}  
		\caption{Efficient repeated transformations with delta-\NDT{} offer opportunities for multi-system/engine designs.}
 \label{fig:neoDB:exp_multiengine}
	\end{center} 
\end{figure}

\begin{figure}[b]
	\begin{center}
        \includegraphics[width=\columnwidth]{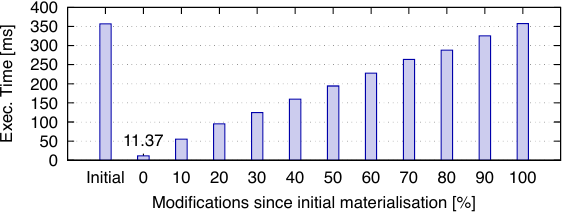}  
\caption{Delta-\NDT{} have consistent and predictable transformation costs.}
	\label{fig:neoDB:exp_delta}
	\end{center} 
\end{figure}

We now demonstrate the efficiency of $\Delta$-\NDT{} (Fig. \ref{fig:neoDB:exp_multiengine} and \ref{fig:neoDB:exp_delta}). A repeated Q6 execution with \NDT{}, after an initial materialization, outperforms PostgreSQL (Fig. \ref{fig:neoDB:exp_multiengine}). Noticeably, the computation time of $\Delta$-\NDT{} increases linearly with the number of modifications (Fig. \ref{fig:neoDB:exp_delta}). With few modifications (e.g., 10\%) $\Delta$-\NDT{} is fast, indicating lower frequent transformation cost, while the $\Delta$-\NDT{} construction costs over a fully updated dataset are as high as the initial ones, indicating the independence of the version chain length.

Another noteworthy observation is that delta-\NDT{} are equally applicable to both multi-engine and multi-systems settings.

\noindent\underline{\textsf{Insight:}} 
\NDT{} offer a potential for reuse, which significantly reducing the cost of full re- transformation. In the presence of OLTP modifications, delta-\NDT{} allows for incrementally transforming only the modified records, efficiently leveraging prior materializations.

\section{Conclusions and Future Directions}
\label{sect:conclusions}
In this paper we propose an approach for performing data (layout) transformations as a near-data processing operation in smart storage. The main idea is to perform operations that mandate heavy data movement, like the transformation, inside smart storage, to leverage its high internal bandwidth and reasonable cost/performance ratio. Besides the opportunities, we see some future research directions.

\paragraph{Poly-Model Smart Storage.} By deploying multiple data transformations as \NDT{}, smart storage can "speak" different data layouts to different types of storage engines. While general data transformations may be too compute-intensive for smart storage, data layout transformations such as KV-to-row (MyRocks), BinaryJASON-to-row (MongoDB) besides row-to-column yield leaner and faster multi-engine systems.

\paragraph{Interfaces.} The applicability of \NDT{} to multi-system settings, calls for novel interfaces to smart/computational storage \cite{SNIA:CompStorage}. We proposed NDBC that eliminates the overhead of wire protocols \cite{Muehleisen:WireProtocol:VLDB:2017}. Further research is needed in the field of SQL (e.g., ADBC, Arrow Flight-SQL) interfaces to smart storage, but also file interfaces as efficient alternatives to \texttt{mmap} \cite{crotty22-mmap}.

\paragraph{Reuse.} We described the concept of reuse for \NDT{}. However, much of its potential on smart storage is in the field of data analytics and machine learning pipelines. To this end deeper integration in the Python tool eco-system is necessary.  Another direction is the application of lightweight compression schemes, e.g., dictionary or run-length encodings during materialization. This will increase the potential for reuse and the applicability in multi-engine settings.

The current software-\NDT{} approach provides remarkable flexibility and programming ease. Nevertheless, when it comes to workhorse algorithms (e.g. compression), data structures, and layouts, there is a promising opportunity for substantial performance and efficiency improvements through hardware/software NDP and the integration of "hardened" accelerators.

\noindent{\bf Acknowledgments.} 
This work has been partially supported by \emph{DFG  neoDBMS -- 419942270}.

\balance

\bibliographystyle{abbrv}
\bibliography{reference}

\end{document}